\documentclass[twocolumn]{aastex63}

\usepackage[T1]{fontenc}
\usepackage{ae,aecompl}
\usepackage{graphicx}
\usepackage{amsmath}
\usepackage{amssymb}
\usepackage{supertabular}
\usepackage{color}
\usepackage{longtable}
\usepackage{booktabs}

\received{}
\revised{}
\accepted{}
\submitjournal{ApJ}
\shorttitle{Analyses on GRB plateaus and flares}
\shortauthors{Yi, Du, \& Liu}

\begin{document}

\title{Statistical analyses on the energies of X-ray plateaus and flares in gamma-ray bursts}

\correspondingauthor{Tong Liu}
\email{tongliu@xmu.edu.cn}

\author[0000-0003-0672-5646]{Shuang-Xi Yi}
\affiliation{School of Physics and Physical Engineering, Qufu Normal University, Qufu, Shandong 273165, China}

\author{Mei Du}
\affiliation{School of Physics and Physical Engineering, Qufu Normal University, Qufu, Shandong 273165, China}

\author[0000-0001-8678-6291]{Tong Liu}
\affiliation{Department of Astronomy, Xiamen University, Xiamen, Fujian 361005, China}

\begin{abstract}
Distinct X-ray plateau and flare phases have been observed in the afterglows of gamma-ray bursts (GRBs), and most of them should be related to central engine activities. In this paper, we collect 174 GRBs with X-ray plateau phases and 106 GRBs with X-ray flares. There are 51 GRBs that overlap in the two selected samples. We analyze the distributions of the proportions of the plateau energy $E_{\rm plateau}$ and the flare energy $E_{\rm flare}$ relative to the isotropic prompt emission energy $E_{\rm \gamma,iso}$. The results indicate that they well meet the Gaussian distributions and the medians of the logarithmic ratios are $\sim -0.96$ and $-1.39$ in the two cases. Moreover, strong positive correlations between $E_{\rm plateau}$ (or $E_{\rm flare}$) and $E_{\rm \gamma,iso}$ with slopes of $\sim 0.95$ (or $\sim0.80$) are presented. For the overlapping sample, the slope is $\sim 0.80$. We argue that most of X-ray plateaus and flares might have the same physical origin but appear with different features because of the different circumstances and radiation mechanisms. We also test the applicabilities of two models, i.e., black holes surrounded by fractured hyperaccretion disks and millisecond magnetars, on the origins of X-ray plateaus and flares.
\end{abstract}


\section{introduction}

Gamma-ray bursts (GRBs) are the most powerful explosions in the universe. According to their duration, GRBs are divided into two groups: long-duration GRBs (LGRBs, $T_{90}> 2~\rm s$) and short-duration GRBs (SGRBs, $T_{90}<2~\rm s$). It is widely believed that they originate from massive collapsars \citep[e.g.,][]{1993ApJ...405..273W,1999ApJ...524..262M} and binary compact object mergers \citep[e.g.,][]{1986ApJ...308L..43P,1989Natur.340..126E,1992ApJ...395L..83N}. Currently, GRB central engines are still uncertain, but there are two leading models, i.e., the stellar-mass black hole (BH) hyperaccretion model \citep[e.g.,][]{1993ApJ...405..273W,1999ApJ...518..356P,2001ApJ...557..949N,2017JHEAp..13....1Y,2018ApJ...852...20L,2019JHEAp..22....5L} and for a review see \citet{2017NewAR..79....1L} and the model of a rapidly rotating neutron star (NS) with a strong magnetic field of $\sim 10^{15}~\rm G$, namely, a millisecond magnetar \citep[e.g.,][]{1992Natur.357..472U,1998A&A...333L..87D,1998PhRvL..81.4301D,2001ApJ...552L..35Z}.

Since Neil Gehrels \emph{Swift} satellite was successfully launched in 2004 \citep{2004ApJ...611.1005G}, both LGRBs and SGRBs with remarkable X-ray plateaus and flares have been observed. It is widely believed that the X-ray plateaus and flares might be linked to the central engine activity, which are different from the X-ray afterglows originating from the fireball shell decelerated in the circumburst medium \citep[e.g.,][]{1992MNRAS.258P..41R,1997ApJ...476..232M,1998ApJ...503..314P,1998ApJ...497L..17S,2005MNRAS.363...93Z,2006MNRAS.369..197F,2013ApJ...776..120Y, 2020ApJ...895...94Y, 2020IJMPD..2950043Z}. Approximately half of \emph{Swift} GRBs show remarkable plateaus or flares in their X-ray afterglows \citep[e.g.,][]{2016ApJS..224...20Y,2019ApJS..245....1T}. X-ray plateaus usually appear in the early afterglow phase with a flat segment, and two different kinds of decay are followed, which are named ``internal" and ``ordinary" plateaus with power-law decay indices of $< -3$ and $> -3$, respectively \citep[see e.g.,][]{2006ApJ...642..389N,2006ApJ...642..354Z,2007RSPTA.365.1179O,2007ApJ...669.1115S,2008MNRAS.391L..79D,2010ApJ...722L.215D,2011ApJ...730..135D,2013ApJ...774..157D,2016ApJ...825L..20D,2020ApJ...901...75D} and for a review see \citet{2017NewAR..77...23D}. A similar shallow decay also occurs in the optical light curves \citep[e.g.,][]{2018ApJ...863...50S,2020ApJ...905L..26D} and \emph{Fermi}-LAT GRBs \citep[e.g.,][]{2021ApJS..255...13D}. The typical timescales of the X-ray plateaus in LGRBs and SGRBs are approximately several hundred and several thousand seconds, respectively. Generally, the extra energy injections from the spin-down of a millisecond magnetar \citep[e.g.,][]{1998A&A...333L..87D,1998PhRvL..81.4301D,2001ApJ...552L..35Z,2010MNRAS.409..531R,2012MNRAS.419.1537B,2013MNRAS.431.1745G,2020ApJ...901...75D} or a supramassive fast-rotating quark star \citep[e.g.,][]{2016PhRvD..94h3010L,2018ApJ...854..104H,2020RAA....20...27O} are expected to explain the plateaus. Other models on the jet modes, such as the two-component ejecta \citep[e.g.,][]{2006ApJ...640L.139T,2009ApJ...690L.118Y}, jets with microphysical factor evolutions \citep[e.g.,][]{2006A&A...458....7I,2006MNRAS.369.2059P} or bulk Lorentz factor distributions \citep[e.g.,][]{2007ApJ...665L..93U}, delayed deceleration jets \citep[e.g.,][]{2015ApJ...806..205D}, and off-axis jets \citep[e.g.,][]{2020MNRAS.492.2847B}, have been proposed. Moreover, \citet{2021ApJ...916...71H} proposed that regardless of the type of central engine, the energy injection driven by the long-lasting precessing jets can also produce plateau phases.

The shapes of X-ray flares generally exhibit a rapid rise and a slow decay \citep[e.g.,][]{2005Sci...309.1833B,2007ApJ...671.1921F,2010MNRAS.406.2113C}. Because of the different temporal behaviors and spectral properties of the afterglows, they are thought to have the same physical origin as the prompt emission of GRBs, and parts of them are triggered by the long-lasting activities or restart of the central engines \citep[e.g.,][]{2010MNRAS.406.2149M,2011MNRAS.417.2144M,2016ApJ...832..161M,2018ApJ...858...34M}. Some theoretical models of X-ray flares have been provided, such as accretion of the fragmentation from a rapidly rotating collapsar \citep{2005ApJ...630L.113K}, differential rotation in a post-merger millisecond magnetar \citep[e.g.,][]{2006Sci...311.1127D}, a magnetic switch of the accretion process \citep{2006MNRAS.370L..61P}, fragmentation of an accretion disk \citep[e.g.,][]{2006ApJ...636L..29P}, transition of the accretion modes \citep{2008MNRAS.388L..15L}, radial or vertical outflows caused by the instabilities of a BH hyperaccretion disk \citep{2008ApJ...676..545L,2014ApJ...791...69L}, He-synthesis-driven disk winds \citep{2009ApJ...699L..93L}, precessing jets \citep[e.g.,][]{2010A&A...516A..16L,2014ApJ...781L..19H}, dynamical instability in jets \citep{2011MNRAS.411L..16L}, episodic jets produced by the magnetohydrodynamic process in an accretion disk \citep{2012ApJ...757...56Y}, and magnetic coupling in a BH hyperaccretion disk \citep{2013ApJ...773..142L}.

The characteristics of X-ray plateaus and flares might indicate that the central engines were always active in the whole explosions. In other words, the duration of central engine activity should be much longer than the observed timescale of GRB prompt emission itself \citep[e.g.,][]{2014ApJ...787...66Z,2018ApJ...852...20L}. However, it is also difficult to determine exactly which type of central engine powers these peculiar variabilities.

In this paper, we collect two samples of GRBs with X-ray plateau and flare activity features and obtain isotropic prompt emission energies $E_{\rm \gamma,iso}$, X-ray plateau energies $E_{\rm plateau}$, and isotropic X-ray flare energies $E_{\rm flare}$. Then, we analyze the distributions of the energy proportions and some correlations between $E_{\rm plateau}$ (or $E_{\rm flare}$) and $E_{\rm \gamma,iso}$, which might provide clues and evidence on their physical mechanisms. This paper is organized as follows. The sample selection and data reduction processes are described in Section 2. In Section 3, we analyze the energy correlations. Two possible models are tested in Section 4, and a brief summary is provided in Section 5.

\section{Data}

\begin{figure*}
\resizebox{92mm}{!}{\includegraphics[]{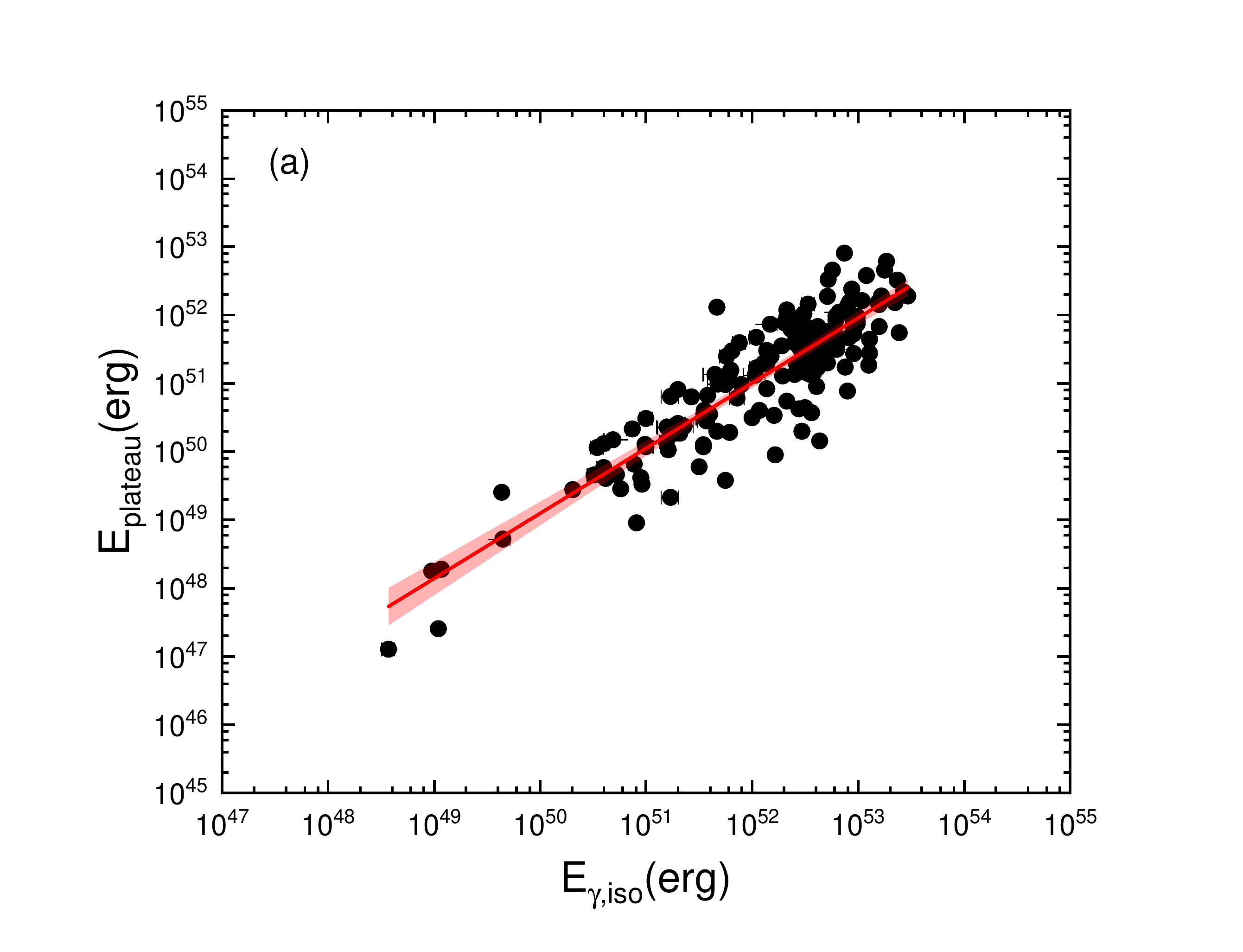}}\resizebox{92mm}{!}{\includegraphics[]{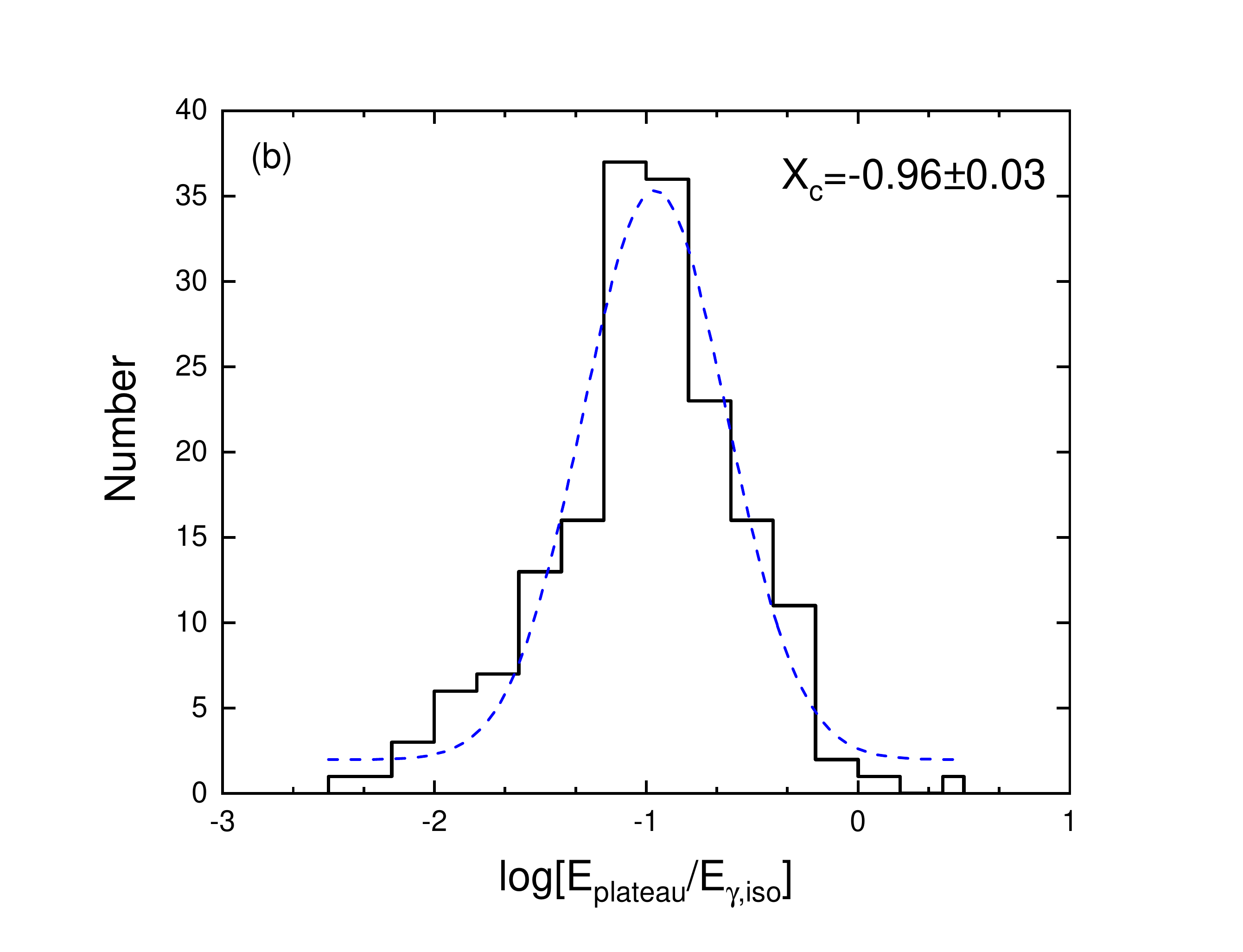}}\\
\caption{\emph{Panel} (a): Correlation of $E_{\rm plateau}-E_{\rm \gamma, iso}$. The red solid line denotes the linear fitting with the slope $\sim 0.95\pm 0.04$, and the light red region represents a 95$\%$ confidence interval. \emph{Panel} (b): Histogram of $\log(E_{\rm plateau}/E_{\rm \gamma, iso})$. The blue dashed line denotes the Gaussian fitting with $X_{\rm c}=-0.96 \pm 0.03$.}
\label{fig:plateau}
\end{figure*}

The observations of \emph{Swift} GRBs promote our knowledge on the taxology and morphology of GRBs, especially the many X-ray plateaus and flares detected in afterglows \citep[e.g.,][]{2006ApJ...642..354Z,2006ApJ...642..389N}. We collected GRBs with plateau and flare features in X-ray afterglows and divided them into two groups. The plateau sample is adopted from \cite{2019ApJS..245....1T}, which includes 174 GRBs observed from May 2005 to August 2018. The screening criteria are extremely strict, and three criteria are applied, i.e., (1) a smoothly broken power-law function is usually used to fit the afterglows with plateaus, and then the power-law index of the plateaus in the range of $\sim -1.0$ to 1.0 is selected; (2) rich data points are required; and (3) the GRB redshift should be given. More details on the data selection can be found in \cite{2019ApJS..245....1T}. In this work, we choose the following GRB parameters: the redshift $z$, the isotropic energy of prompt emission $E_{\rm \gamma, iso}$, and the isotropic plateau energy $E_{\rm plateau}$, which are shown in Table 1. The plateau phase energy is given by $E_{\rm plateau}=T_{\rm a}L_X(1+z)$, where $T_{\rm a}$ is the end time of the plateau phase in the GRB rest frame and $L_{\rm X}$ is the X-ray luminosity at the end time. This correlation was already previously discovered in \citet{2008MNRAS.391L..79D,2010ApJ...722L.215D,2013MNRAS.436...82D,2015MNRAS.451.3898D,2017ApJ...848...88D}. Most GRBs with plateaus belong to LGRBs, and only a small fraction are classified as SGRBs, including GRBs 051221A, 061201, 070809, 090510, 130603B, 140903A, and 150423A.

Furthermore, 106 GRBs with X-ray flares are collected. Most of them are taken from \citet{2016ApJS..224...20Y}, and the others are adopted from \citet{2007ApJ...671.1921F}, \citet{2010MNRAS.406.2113C}, \citet{2011A&A...526A..27B}, and \citet{2015ApJ...807...92Y}. The selected GRBs with X-ray flares are all LGRBs. For GRBs with two or more flares, the sum of these flare energies is considered as the flare energy $E_{\rm flare}$, which can be obtained by $E_{\rm flare}=4\pi D^2_{\rm L}S_{\rm F}/(1+z)$, where $D_{\rm L}$ and $S_{\rm F}$ are the luminosity distance and the fluence of X-ray flares, respectively. Meanwhile, the corresponding $E_{\rm \gamma,iso}$ can be obtained from \citet{2016ApJS..227....7L}, \citet{2018ApJ...852...53R}, \citet{2018JHEAp..18...21W}, \citet{2019ApJS..245....1T}, \citet{2019ApJ...887...13F}, \citet{2020MNRAS.492.1919M}, \citet{2020ApJS..248...21H}, and \citet{2021ApJ...908..242D}. The X-ray flare data, including the redshift $z$, the isotropic energy of prompt emission $E_{\rm \gamma, iso}$, and the total flare energy $E_{\rm flare}$, are listed in Table 2.

Interestingly, 51 GRBs overlapped in the two samples, which means that both plateaus and flares appeared in their X-ray afterglows. The overlapping data are labeled in boldface in Tables 1 and 2.

\section{Statistical analyses}

\begin{figure*}
\resizebox{92mm}{!}{\includegraphics[]{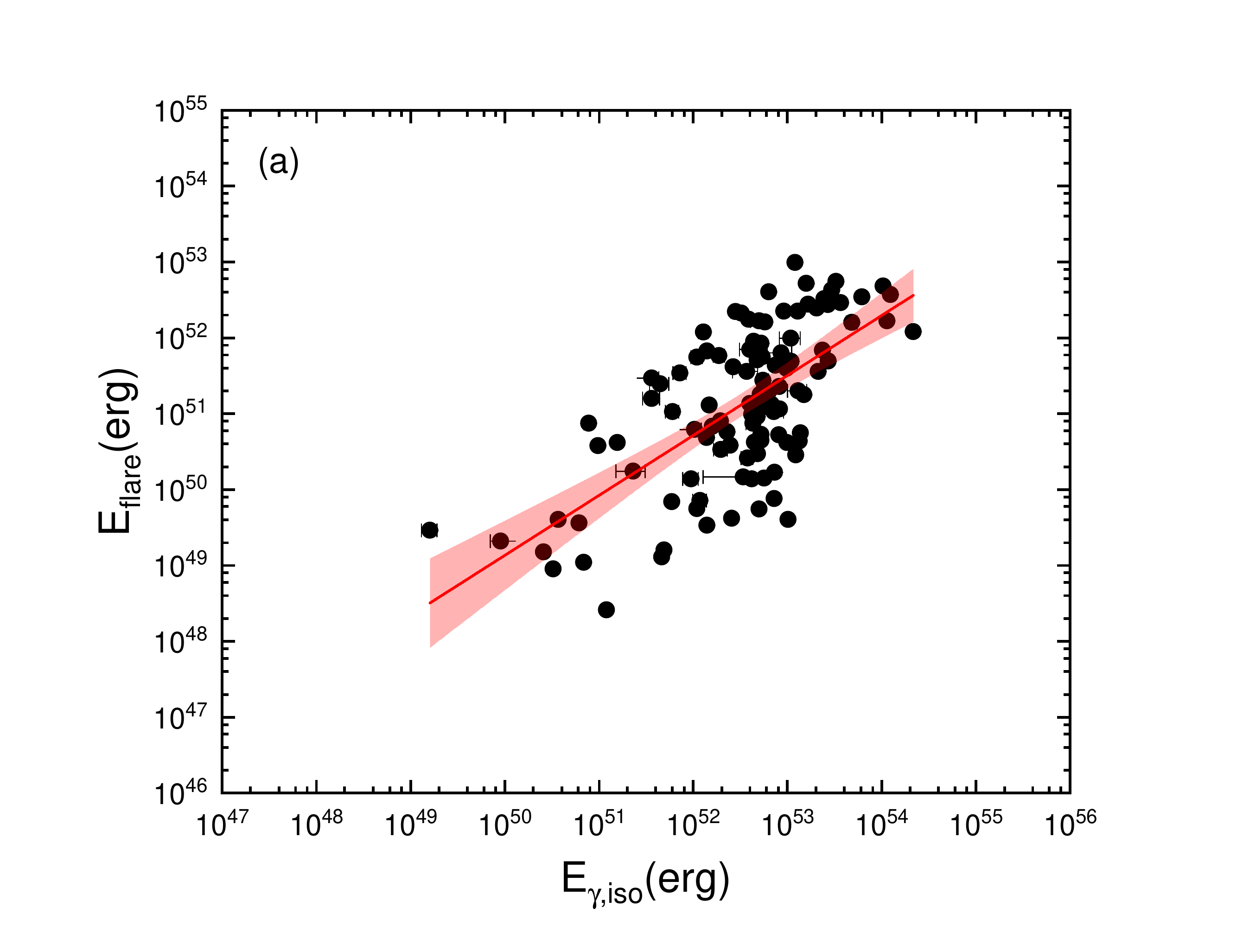}}\resizebox{92mm}{!}{\includegraphics[]{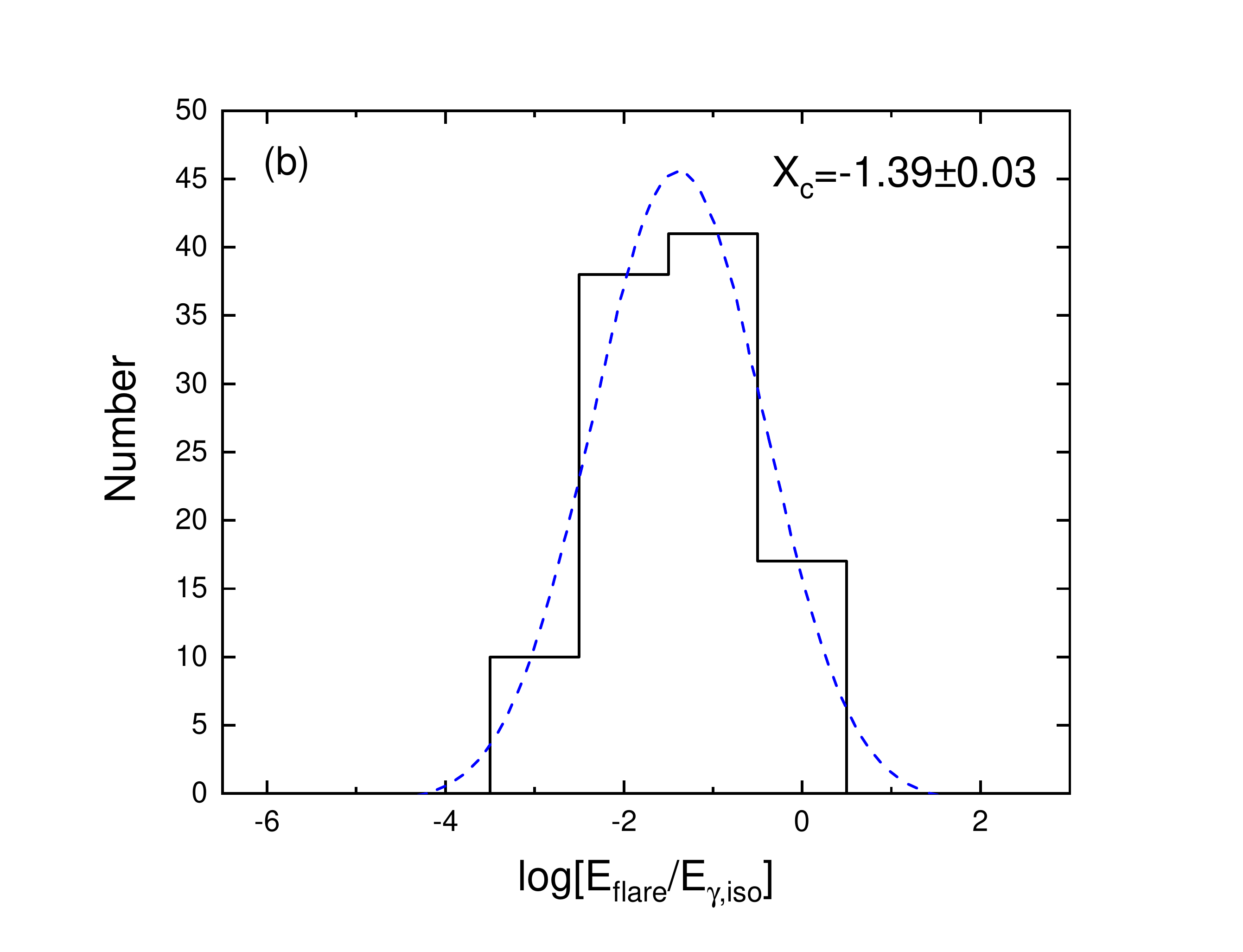}}\\
\caption{\emph{Panel} (a): Similar to Figure 1(a) but for X-ray flares with a slope of $\sim 0.80\pm 0.09$. \emph{Panel} (b): Similar to Figure 1(b) but $X_{\rm c}=-1.37 \pm 0.02$.}
\label{fig:flare}
\end{figure*}

Since most X-ray plateaus and flares are closely related to the activities of the GRB central engines, we speculate that they might have similar statistical properties. Then, we study these possibilities by analyzing the statistical relations between $E_{\rm plateau}$ (or $E_{\rm flare}$) and $E_{\rm \gamma,iso}$, as well as the ratio of $E_{\rm plateau}$ (or $E_{\rm flare}$) to $E_{\rm \gamma,iso}$. The first time $E_{\rm \gamma, iso}$ v.s. $L_X$ was proposed by \citet{2011MNRAS.418.2202D}, later updated in \citet{2015ApJ...800...31D}. Also $E_{\rm \gamma, iso}$ v.s. $T_a$ has been investigated, where $L_X-T_a$ are anti-correlated \citep{2008MNRAS.391L..79D}.

Figures 1(a) and 1(b) show the correlation between $E_{\rm plateau}$ and $E_{\rm \gamma,iso}$ and the distributions of $\log(E_{\rm plateau}/E_{\rm \gamma,iso})$, respectively. Obviously, the positive correlation with a slope of $\sim 0.95\pm 0.04$ is very strong, and the median of the Gaussian distribution on $\log (E_{\rm plateau}/E_{\rm \gamma,iso})$ is approximately $-0.96 \pm 0.03$, i.e., the typical plateau energy $E_{\rm plateau}$ is approximately a tenth of the typical energy of prompt emission $E_{\rm \gamma,iso}$.

Figures 2(a) and 2(b) display a similar correlation and contribution but for the isotropic flare energy $E_{\rm flare}$. The positive correlation between $E_{\rm flare}$ and $E_{\rm \gamma,iso}$ with a slope of $\sim 0.80\pm 0.09$ is slightly looser than the correction in Figure 1(a). We consider that this issue might be resulted in some of X-ray flares related to the external origins \citep[e.g.,][]{2010MNRAS.406.2149M,2011MNRAS.417.2144M,2016ApJ...832..161M}. The median of the Gaussian distribution on $\log(E_{\rm flare}/E_{\rm \gamma,iso})$ is approximately $-1.39 \pm 0.03$. Interestingly, the behaviors of the above correlation and distribution on the plateaus are similar to those on the flares.

In addition, Figure 3 shows the correlations and distributions of 51 GRBs with both X-ray plateaus and flares. The correlations of $E_{\rm plateau}-E_{\rm \gamma,iso}$, $E_{\rm flare}-E_{\rm\gamma,iso}$, and $(E_{\rm plateau}+E_{\rm flare})-E_{\rm\gamma, iso}$ for 51 GRBs have almost the same tendency with slopes of $\sim 0.85\pm0.09$, $0.76\pm0.14$, and $0.82\pm 0.08$, respectively. The Gaussian distributions for $\log(E_{\rm plateau}/E_{\rm\gamma,iso})$ and $\log(E_{\rm flare}/E_{\rm \gamma,iso})$ of the overlapping GRBs show similar Gaussian distributions, and the central values are approximately $-1.10 \pm 0.06$ and $-0.98 \pm 0.05$, respectively. As shown in Figure 3(f), the distribution of $\log [(E_{\rm plateau}+E_{\rm flare})/E_{\rm \gamma,iso}]$ shows a bimodal distribution, which is caused by the amplifying dispersions once the values of $E_{\rm plateau}$ and $E_{\rm flare}$ mixed together, but the correlation of ($E_{\rm plateau}+E_{\rm flare})-E_{\rm \gamma,iso}$ still seems tight with a slope of $\sim 0.82 \pm 0.08$. All of the best linear fitting results in the left panels of Figures 1-3 are shown in Table 3.

Although the shapes and features of X-ray plateaus are definitely different from those of flares, the above statistical analyses indicate that both X-ray plateaus and flares have similar ratios of energy release from the central engine and similar correlations to the isotropic energy of prompt emission. Thus, we argue that most of X-ray plateaus and flares should have the same physical origin, i.e., a new round of the energy releases from the active central engines, and the different features might be caused by different circumstances and radiation mechanisms.

\section{Testing of central engine models}

\begin{figure*}[ht!]
\center
\resizebox{92mm}{!}{\includegraphics[]{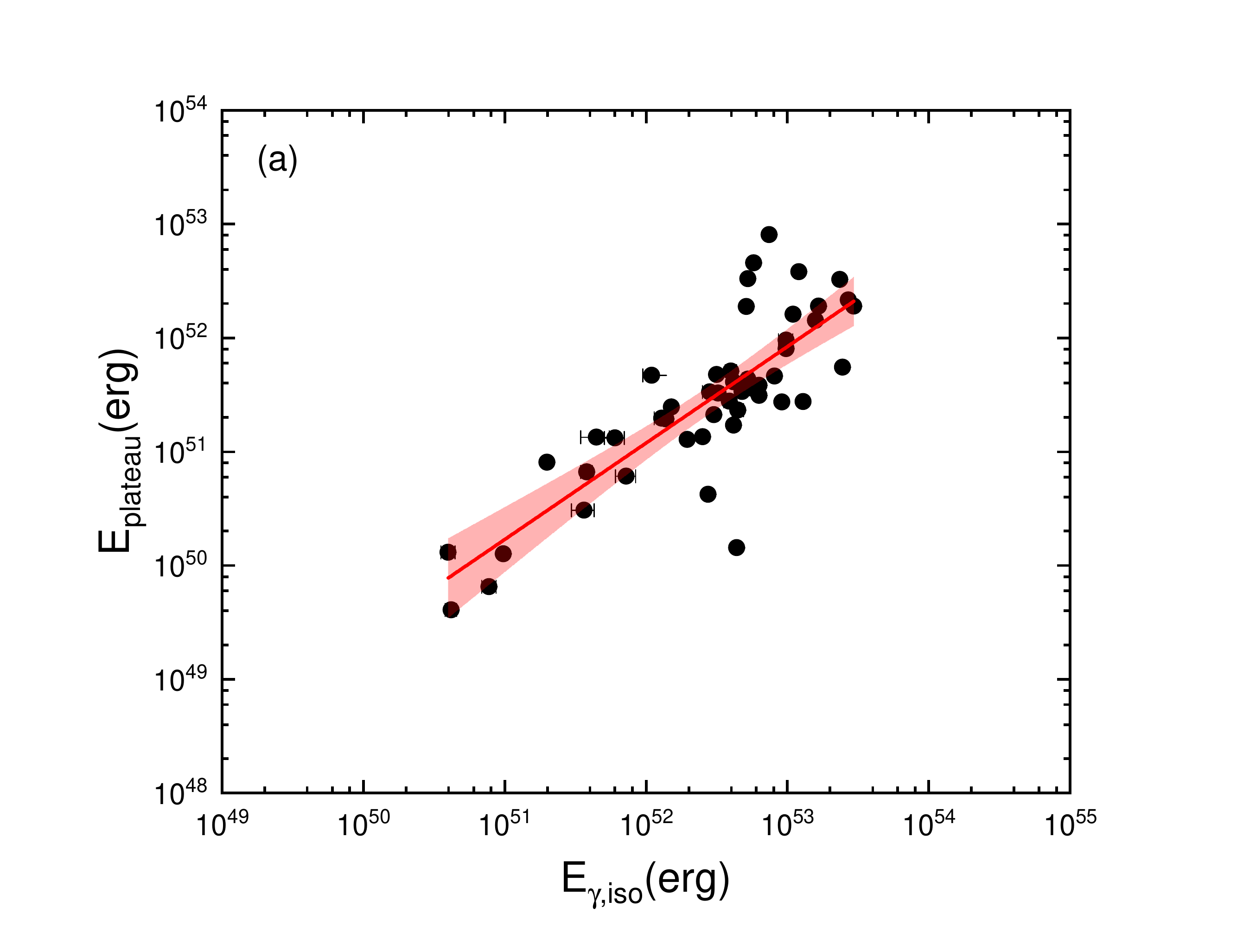}}\resizebox{92mm}{!}{\includegraphics[]{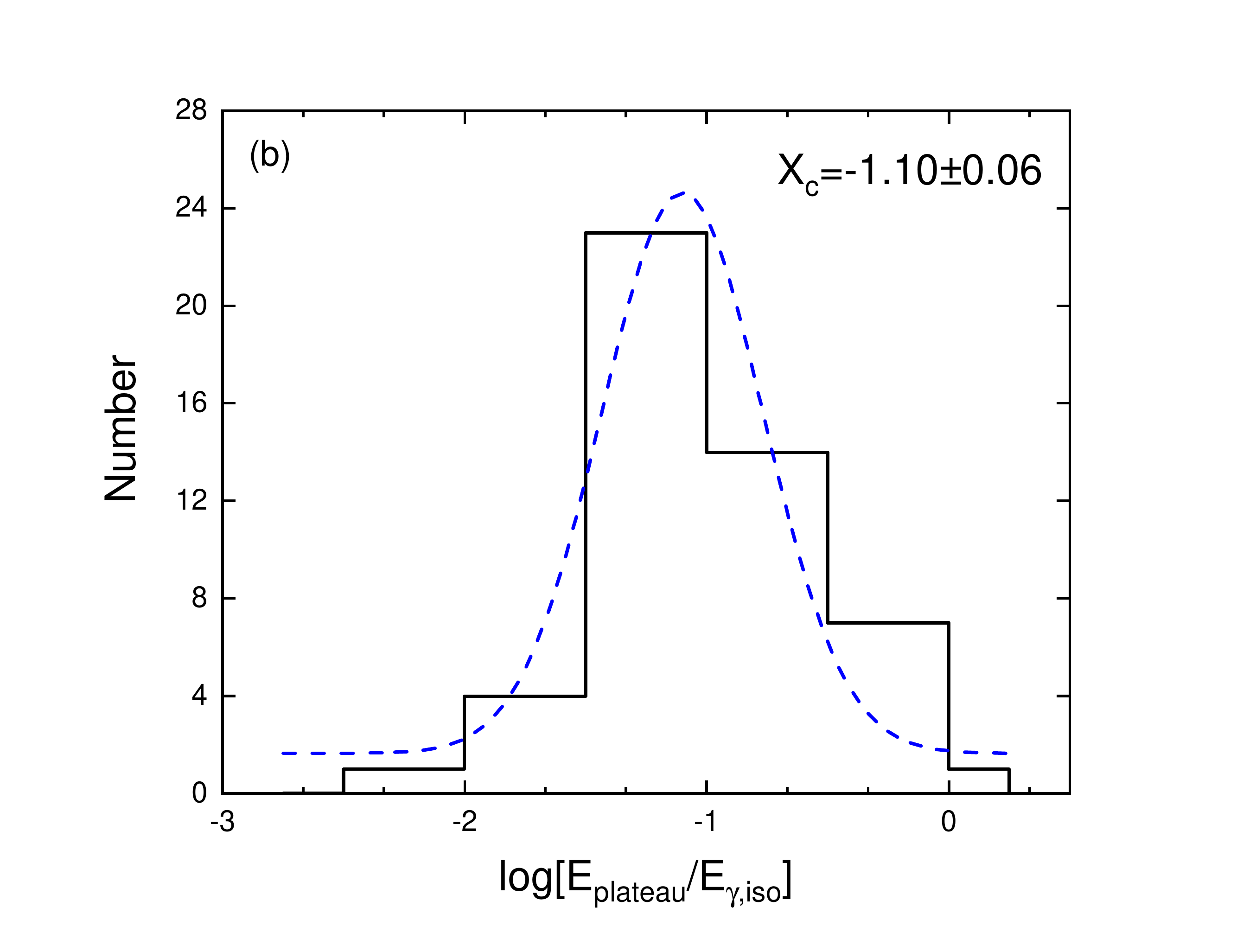}}\\
\resizebox{92mm}{!}{\includegraphics[]{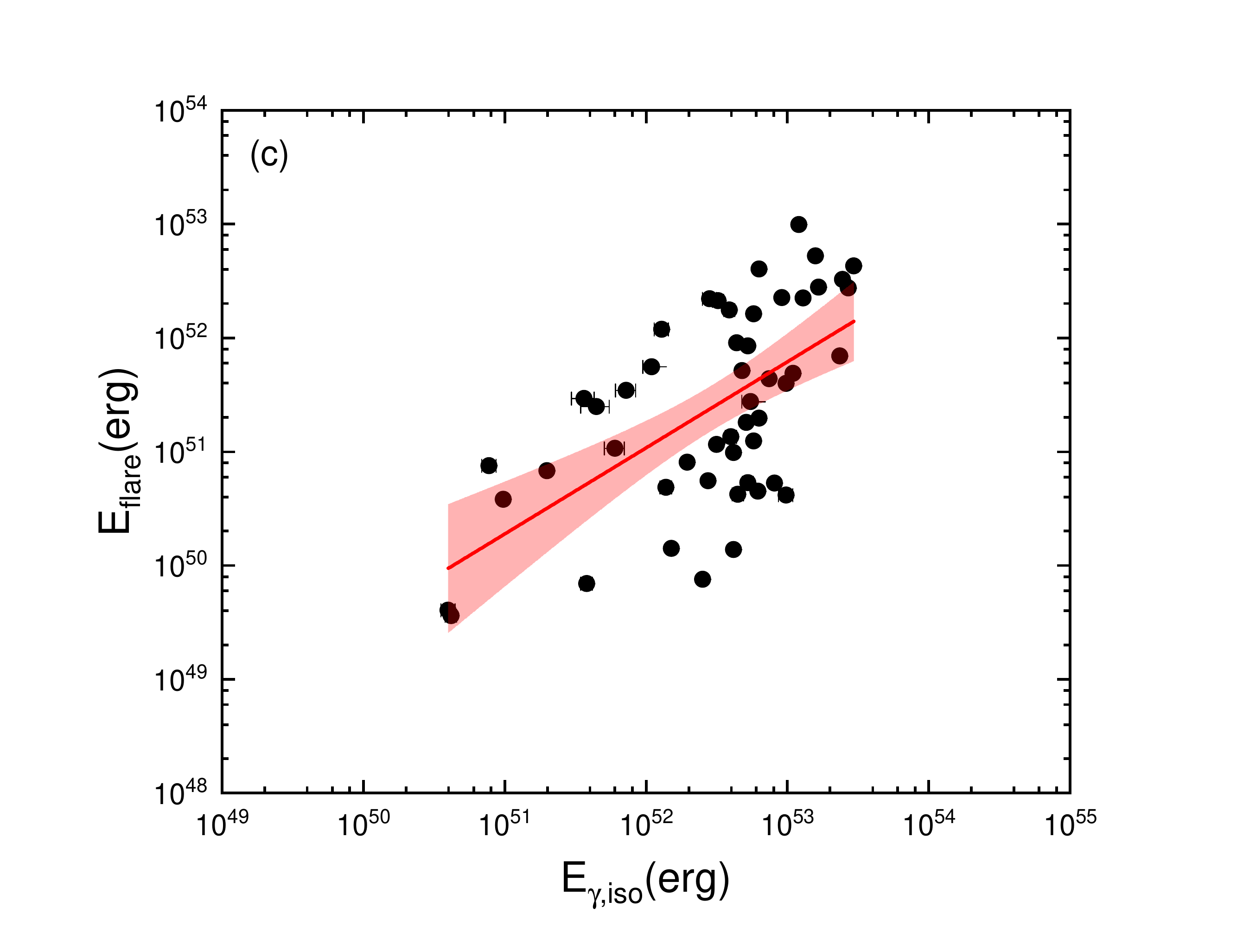}}\resizebox{92mm}{!}{\includegraphics[]{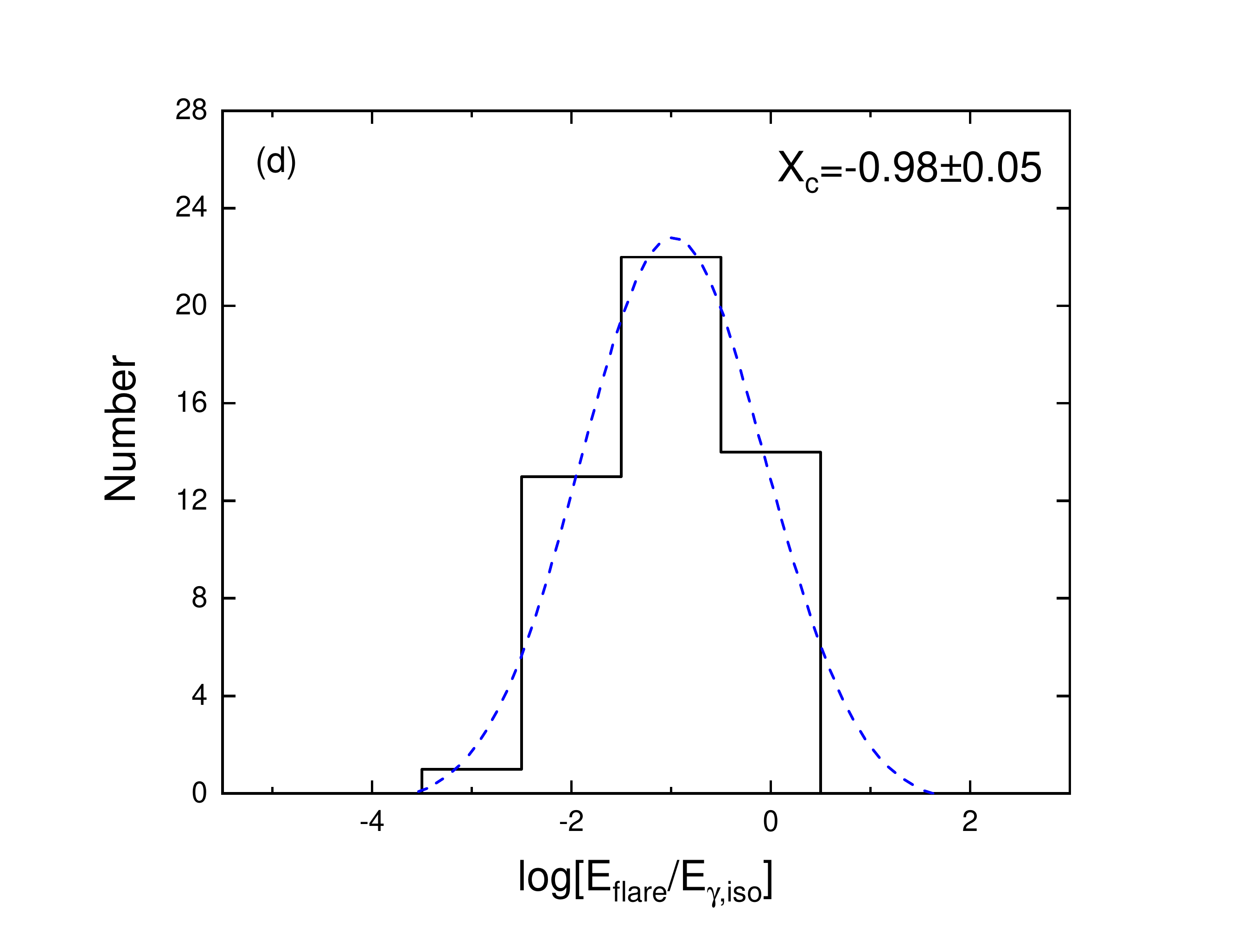}}\\
\resizebox{92mm}{!}{\includegraphics[]{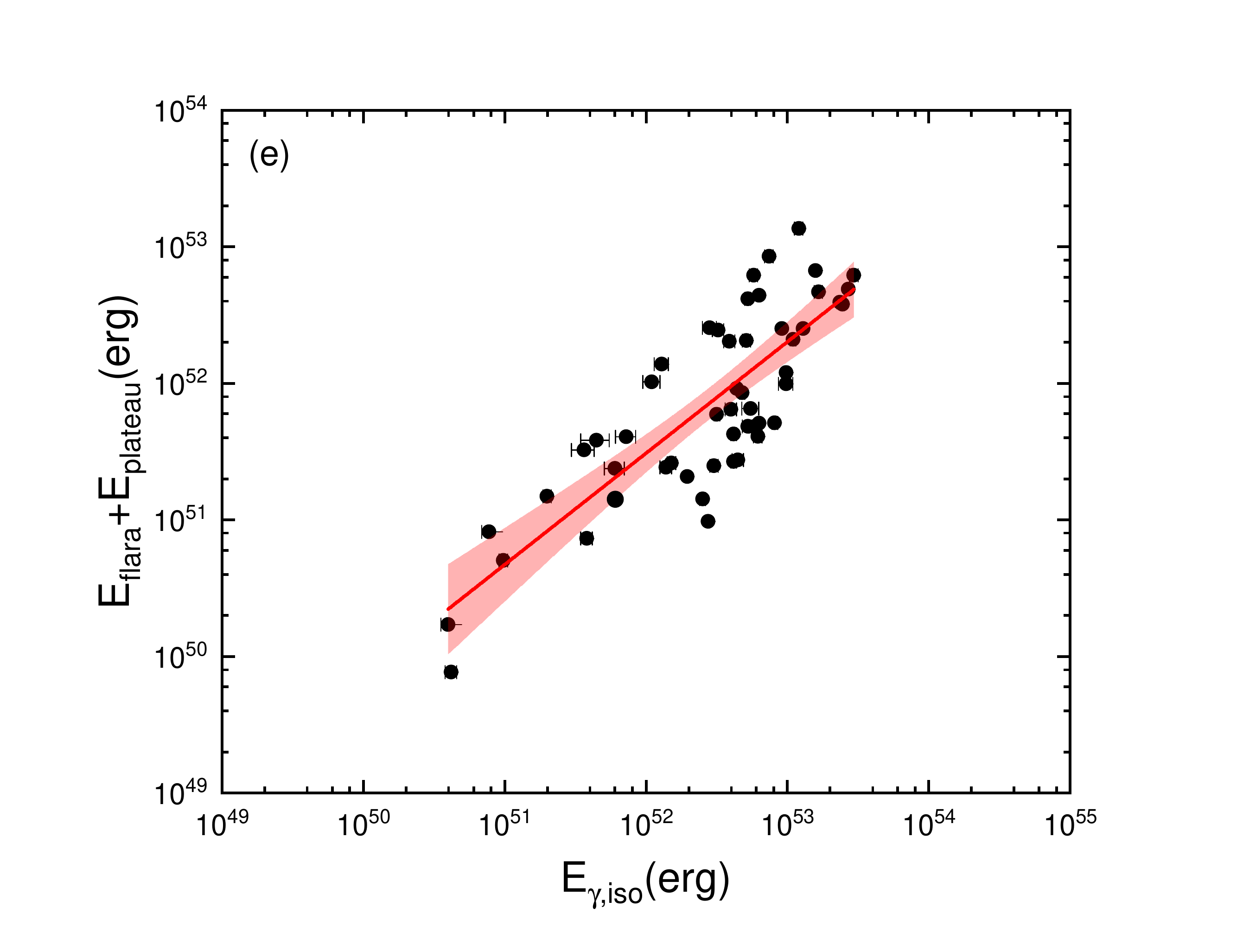}}\resizebox{92mm}{!}{\includegraphics[]{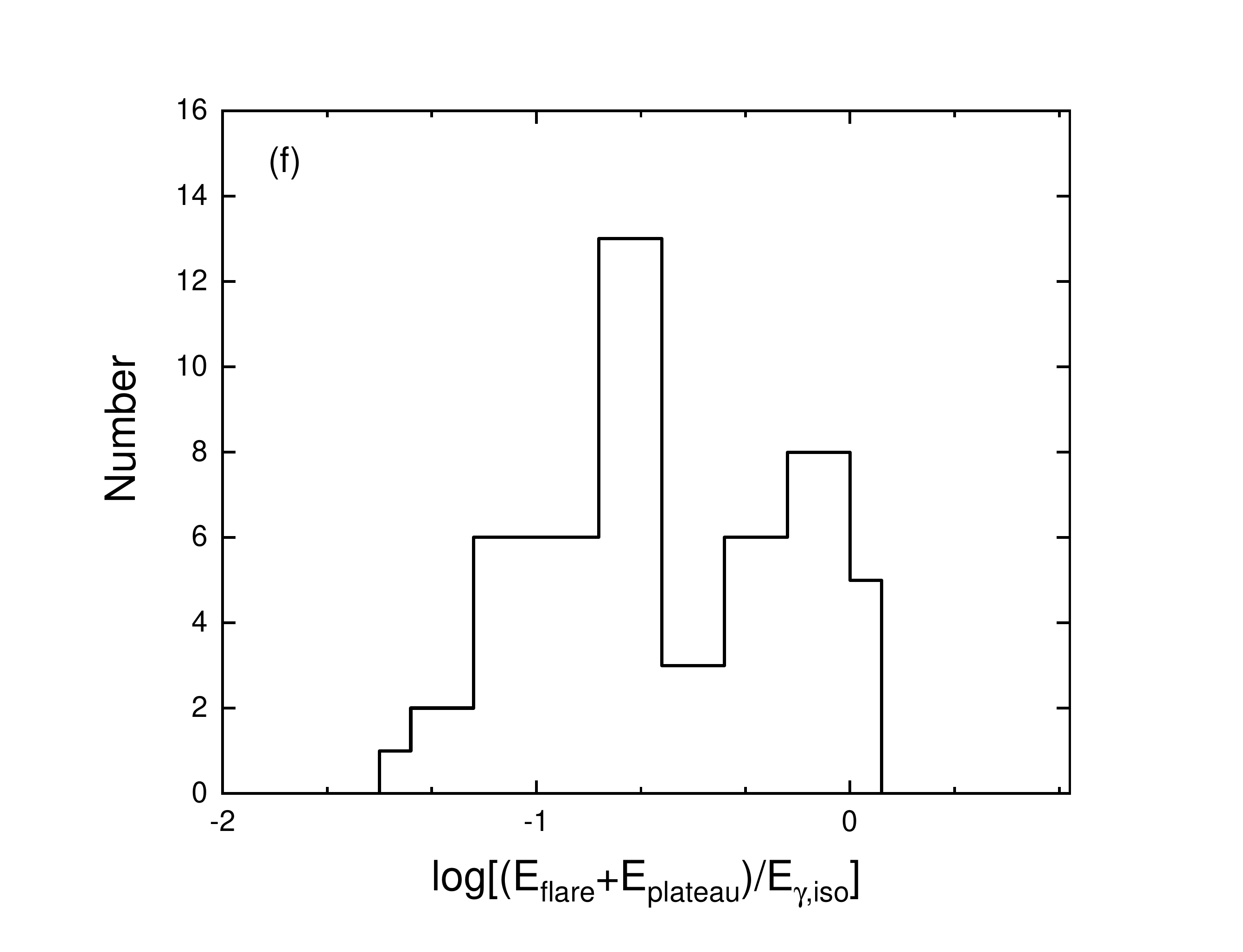}}\\
\caption{\emph{Panels} (a, c, e): Correlations of $E_{\rm plateau}-E_{\rm\gamma, iso}$, $E_{\rm flare}-E_{\rm \gamma, iso}$, and $(E_{\rm plateau}+E_{\rm flare})-E_{\rm\gamma, iso}$ for 51 GRBs with both flares and plateaus. The red solid lines correspond to the linear fitting with slopes of $\sim 0.85\pm 0.09$, $0.76\pm 0.14$, and $0.82\pm 0.08$. \emph{Panels} (b, d, f): Histogram of $\log(E_{\rm plateau}/E_{\rm \gamma, iso})$, $\log(E_{\rm flare}/E_{\rm\gamma, iso})$, and $\log[(E_{\rm plateau}+E_{\rm flare})/E_{\rm\gamma, iso}]$ for these 51 GRBs. The blue dashed lines are the Gaussian fitting lines with $X_{\rm c}=-1.10 \pm 0.06$ and $-0.98 \pm 0.05$ in \emph{Panels} (b) and (d), respectively.}
\label{fig:flare+plateau}
\end{figure*}

It is difficult to determine the exact type of central engine based on the GRB observations. Since a stellar-mass BH or an NS should be born in the center of a massive collapsar and compact object merger, the most plausible candidates of the GRB central engine models could be confirmed, i.e., BH hyperaccretion disk and millisecond magnetar. According to the above statistical results, most X-ray plateaus and flares are triggered by central engine activities, so here, we test two popular models to roughly discuss their possible origins and estimate the main influencing factors in the scenarios of BH hyperaccretion and magnetars.

\subsection{BH hyperaccretion}

\citet{2006ApJ...636L..29P} proposed that GRB X-ray flares could be powered by fractured accretion disks. If gravitational instability exists in the accretion disk and the local cooling of the disk is rapid \citep[e.g.,][]{2000ApJ...544L..91N,2014ApJ...791...69L}, some fragments might be formed in the outer part of the accretion disk. Then, the fragments falling into the BH with a relatively slow in-spiral will lead to large-amplitude changes in the accretion rate; thus, X-ray flares will appear after prompt emission. In this framework, the released energy of each flare is roughly proportional to the mass of each fragment in the outer region of the disk. We consider that this mechanism is also suitable for energy-ejection-driven X-ray plateaus. In the case of the rapid cooling, episodic energy injection can also produce the shapes of plateaus, which has been verified \citep{2021ApJ...916...71H}; If the slow cooling appears in the disk, it will lead to the redistribution of the disk materials and then maintain a long-lived accretion process \citep[e.g.,][]{2014ApJ...791...69L}.

The ratio of the formed fragment (or the extended disk) mass $M_{\rm frag}$ to the central BH mass $M_{\rm BH}$ can be expressed as \citep[e.g.,][]{1996ApJ...460..832T,2006ApJ...636L..29P}
\begin{equation}
\frac{M_{\rm frag}}{M_{\rm BH}} \simeq \left(\frac{H}R\right)^2\alpha^{1/2}
\end{equation}
where $H$, $R$, and $\alpha$ are the half-thickness of the disk, the distance from the central object to the fragments, and the viscosity parameter, respectively \citep{1973A&A....24..337S,2008bhad.book.....K}.

For the extremely geometrically thick disk, $H \sim R$, and the typical value of $M_{\rm BH}$ is $\sim 3~M_\odot$. Therefore, we obtain $M_{\rm frag} \sim 3 \alpha^{1/2}~M_\odot$. As the main mechanisms of energy release of GRBs, the neutrino annihilation process \citep[e.g.,][]{2017NewAR..79....1L} and Blandford-Znajek mechanism \citep{1977MNRAS.179..433B,2000ApJ...536..416L} ask the typical accretion mass $\sim 3$ and $1~M_\odot$ for the prompt emission of LGRBs, respectively \citep{2015ApJS..218...12L}. Considering the above results $E_{\rm plateau}/E_{\rm \gamma,iso}$ (or $E_{\rm flare}/E_{\rm \gamma,iso}$) $\sim 0.1$, we obtain $\alpha\sim 0.01$, which satisfies the requirements on the cooling timescale and gravitational instability \citep{2006ApJ...636L..29P}.

\subsection{Magnetars}

In the magnetar scenario, the energy injection can be driven by the spin-down of magnetars to power X-ray plateaus. The total rotation energy can be given by
\begin{equation}
E_{\rm rot}=\frac{1}{2}I\Omega_{0}\simeq 2\times 10^{52} M_{\rm NS,1.4} R_{\rm NS,6}^{2} P_{0,-3}^{-2} ~{\rm erg},
\end{equation}
where $I$ is the moment of inertia, $\Omega_{0}=2\pi/P_0$ is the initial angular frequency, and $M_{\rm NS,1.4}=M/(1.4~M_{\odot})$, $R_{\rm NS,6}=R_{\rm NS}/(10^6 ~\rm cm)$, and $P_{0,-3}=P_0/(10^{-3}~\rm s)$ are the dimensionless NS mass, radius, and rotation period, respectively. The LT relation of X-ray plateaus \citep[e.g.,][]{2008MNRAS.391L..79D} can be explained by the magnetar model \citep[e.g.,][]{2014MNRAS.443.1779R,2015ApJ...813...92R,2018ApJ...869..155S}. Of course the winds driven by the rotations could power the X-ray flares \citep[e.g.,][]{2006Sci...311.1127D,2011MNRAS.413.2031M,2016ApJ...832..161M}.

However, the mechanism of the energy release in the GRB prompt emission phase remains uncertain. \citet{1998ApJ...505L.113K}, \citet{2000ApJ...542..243R}, and \citet{2006Sci...311.1127D} pointed out that a newborn NS may be unstable, the core and the exterior have different spin speeds, and differential rotation leads to an internal poloidal magnetic field windup. Then, the produced toroidal fields are so strong that the fields float up and break through the surface. An outburst event caused by the magnetic reconnection occurred next, and the burst energy $E_{\rm b}$ of each eruption may be estimated as \citep[e.g.,][]{1998ApJ...505L.113K}
\begin{equation}
E_{\rm b} \simeq 6 \times 10^{51} \frac{V_{\rm B}}{V_*} ~\rm erg,
\end{equation}
where $V_{\rm B}$ and $V_*$ are the volume of the toroid and the NS, respectively. For a GRB event, the sum of frequent energy released in the GRB prompt emission phase can exceed $\sim 10^{53}$ erg \citep[e.g.,][]{1998ApJ...505L.113K}. Then, the ratio of the total or parts of spin-down energy to the energy of differential rotation is easily taken as $\sim 0.1$, which conforms to the statistical results.

In conclusion, the BH hyperaccretion and magnetar models are both suitable to explain the observations of GRB X-ray plateaus and flares. In other words, the verification of the central engines is practicable on their uniform origins, but there is no evidence to constrain and distinguish these models. Moreover, besides the above mentioned BH or NS related central engines, only in the jet modes, the precessing or off-axis jets launched from the whichever central engines are considered as the possible origins of both X-ray plateaus and flares \citep[e.g.,][]{2010A&A...516A..16L,2021ApJ...916...71H}.

\section{Summary}

In this paper, two samples, including 174 GRBs with plateau phases and 106 GRBs with X-ray flares, are exhibited, of which 51 GRBs overlap in the two samples. One can note three tightly positive correlations with similar slopes for $E_{\rm plateau}-E_{\rm\gamma,iso}$, $E_{\rm flare}-E_{\rm \gamma,iso}$, and $(E_{\rm plateau}+E_{\rm flare})-E_{\rm \gamma,iso}$. The distributions of $E_{\rm plateau}/E_{\rm \gamma,iso}$ and $E_{\rm flare}/E_{\rm \gamma,iso}$ can be fitted with a Gaussian distribution very well, and their median is approximately $\sim 0.1$, especially for the overlapping GRB sample. The $E_{\rm plateau}-E_{\rm\gamma,iso}$ correlation still holds for more cases confirming the result of \citet{2011MNRAS.418.2202D}. We argue that both X-ray plateaus and flares might have the same physical origin, but appear with different features in GRB afterglows. We also discuss the applicabilities of two popular models of GRB central engines, i.e., BHs surrounded by fractured hyperaccretion disks and millisecond magnetars.

The central engine of GRBs is still a mystery, and many models have been proposed in the literature. Depending only on multiband observations, it is difficult to solve this puzzle, but the multimessenger era might offer the possibility via joint detection of electromagnetic radiation, gravitational waves, and MeV neutrinos related to the central engine and progenitors \citep[e.g.,][]{2017NewAR..79....1L}.

\acknowledgments
We thank Zi-Gao Dai and Shu-Qing Zhong for helpful discussion. This work was supported by the National Natural Science Foundation of China under grants U2038106, 11822304, and 12173031.

\clearpage

\begin{longtable*}[ht]{cccc||cccc}
  \caption{Data of GRBs with X-ray plateaus.}\\
  \toprule
    GRB   & $z$  & $E_{\rm \gamma,iso}~(10^{51}~\rm erg)$ & $E_{\rm plateau}~(10^{50}~\rm erg)$& GRB  & $z$   & $E_{\rm \gamma,iso}~(10^{51}~\rm erg)$ & $E_{\rm plateau}~(10^{50}~\rm erg)$ \\
  \midrule
    050315 & 1.949 & 30.200$\pm$1.391 & 51.248 &
    050319 & 3.24  & 29.512$\pm$3.398 & 1.983 \\
    050401 & 2.9   & 154.882$\pm$7.133 & 151.728&
    \textbf{050416A} & 0.6535 & 0.417$\pm$0.038 & 0.406 \\
    050505 & 4.27  & 87.096$\pm$6.016 & 240.885&
    \textbf{050730} & 3.97  & 74.131$\pm$5.121 & 806.04 \\
    050801 & 1.38  & 1.549$\pm$0.214 & 1.401 &
    \textbf{050802} & 1.71  & 15.136$\pm$1.046 & 24.715 \\
    \textbf{050803} & 0.422 & 0.977$\pm$0.068 & 1.267 &
    \textbf{050814} & 5.3   & 97.724$\pm$11.251 & 95.354 \\
    050824 & 0.83  & 0.490$\pm$0.090 & 1.487 &
    050922C & 2.199 & 19.055$\pm$0.439 & 35.076 \\
    \textbf{051016B} & 0.9364 & 0.398$\pm$0.046 & 1.309 &
    051109A & 2.346 & 28.840$\pm$3.320 & 66.761 \\
    051109B & 0.08  & 0.004$\pm$0.001 & 0.001 &
    051221A & 0.547 & 0.912$\pm$0.021 & 0.331 \\
    \textbf{060108} & 2.03  & 3.802$\pm$0.350 & 6.629 &
    \textbf{060115} & 3.53  & 44.668$\pm$4.114 & 23.233 \\
    060116 & 4     & 75.858$\pm$6.987 & 17.337 &
    060202 & 0.78  & 3.467$\pm$0.240 & 1.176 \\
    060206 & 4.05  & 26.915$\pm$1.240 & 66.572 &
    \textbf{060210} & 3.91  & 234.423$\pm$10.796 & 324.4 \\
    060502A & 1.51  & 13.804$\pm$0.636 & 8.311 &
    \textbf{060522} & 5.11  & 52.481$\pm$4.834 & 43.255 \\
    \textbf{060526} & 3.21  & 28.184$\pm$3.245 & 33.441 &
    \textbf{060604} & 2.1357 & 4.467$\pm$1.029 & 13.376 \\
    060605 & 3.8   & 20.417$\pm$2.351 & 76.075 &
    \textbf{060607A} & 3.082 & 52.481$\pm$2.417 & 331.797\\
    060614 & 0.13  & 0.813$\pm$0.019 & 0.09 &
    \textbf{060707} & 3.43  & 39.811$\pm$3.667 & 50.863 \\
    060708 & 2.3   & 6.310$\pm$0.436 & 15.435 &
    \textbf{060714} & 2.71  & 47.863$\pm$2.204 & 33.836 \\
    \textbf{060729} & 0.54  & 1.995$\pm$0.138 & 8.082 &
    \textbf{060814} & 0.84  & 27.542$\pm$0.634 & 4.215\\
    \textbf{060906} & 3.685 & 61.660$\pm$4.259 & 36.367 &
    \textbf{060908} & 1.8836 & 25.119$\pm$1.157 & 13.488 \\
    060912A & 0.937 & 3.162$\pm$0.146 & 0.599 &
    061021 & 0.3463 & 0.891$\pm$0.021 & 0.416 \\
    \textbf{061121} & 1.314 & 63.096$\pm$1.453 & 31.215 &
    061201 & 0.111 & 0.010$\pm$0.001 & 0.018 \\
    061222A & 2.088 & 85.114$\pm$1.960 & 99.926 &
    070110 & 2.352 & 21.380$\pm$1.477 & 118.933 \\
    \textbf{070129} & 2.3384 & 38.905$\pm$3.583 & 27.768 &
    070208 & 1.165 & 1.622$\pm$0.336 & 2.215 \\
    \textbf{070306} & 1.497 & 31.623$\pm$1.456 & 47.579 &
    070506 & 2.31  & 2.692$\pm$0.310 & 6.307 \\
    070508 & 0.82  & 35.481$\pm$0.817 & 13.492 &
    070529 & 2.4996 & 38.019$\pm$3.502 & 13.932 \\
    070714B & 0.92  & 1.622$\pm$0.187 & 1.055 &
    \textbf{070721B} & 3.626 & 97.724$\pm$4.500 & 80.391 \\
    070809 & 0.22  & 0.012$\pm$0.001 & 0.019 &
    070810A & 2.17  & 7.943$\pm$0.732 & 9.573 \\
    071020 & 2.145 & 25.704$\pm$1.184 & 38.692 &
    \textbf{080310} & 2.43  & 32.359$\pm$2.980 & 32.756 \\
    080430 & 0.767 & 1.905$\pm$0.132 & 1.85 &
    080516 & 3.2   & 5.754$\pm$0.795 & 24.731 \\
    080603B & 2.69  & 39.811$\pm$1.833 & 53.337 &
    080605 & 1.6398 & 91.201$\pm$2.100 & 52.671 \\
    080707 & 1.23  & 2.089$\pm$0.241 & 1.855 &
    080721 & 2.602 & 186.209$\pm$12.863 & 611.707 \\
    \textbf{080810} & 3.35  & 109.648$\pm$5.049 & 161.618 &
    080905B & 2.374 & 23.988$\pm$2.762 & 88.745 \\
    081007 & 0.5295 & 0.525$\pm$0.060 & 0.462 &
    \textbf{081008} & 1.9685 & 41.687$\pm$1.920 & 17.082 \\
    081029 & 3.8479 & 61.660$\pm$5.679 & 96.728 &
    081221 & 2.26  & 223.872$\pm$5.155 & 152.482 \\
    090113 & 1.7493 & 5.888$\pm$0.271 & 10.696 &
    090205 & 4.7   & 7.586$\pm$1.048 & 39.434 \\
    090313 & 3.375 & 33.884$\pm$4.681 & 144.87 &
    \textbf{090407} & 1.4485 & 6.026$\pm$0.971 & 13.149 \\
    090418A & 1.608 & 30.903$\pm$1.423 & 27.945 &
    \textbf{090423} & 8     & 51.286$\pm$3.543 & 188.037 \\
    090510 & 0.903 & 0.741$\pm$0.085 & 2.135 &
    \textbf{090516} & 4.109 & 295.121$\pm$20.386 & 189.817 \\
    090519 & 3.9   & 36.308$\pm$2.508 & 3.717 &
    090529 & 2.625 & 10.715$\pm$2.467 & 13.162 \\
    090530 & 1.266 & 4.677$\pm$0.431 & 1.974 &
    090618 & 0.54  & 79.433  & 7.718 \\
    090927 & 1.37  & 1.000$\pm$0.138 & 3.053 &
    091018 & 0.971 & 3.548$\pm$0.245 & 4.024 \\
    \textbf{091029} & 2.752 & 41.687$\pm$1.920 & 41.14 &
    091109A & 3.5   & 40.738$\pm$4.690 & 8.979 \\
    091127 & 0.49  & 5.623$\pm$0.129 & 9.62 &
    091208B & 1.0633 & 10.000$\pm$0.691 & 3.123 \\
    100219A & 4.7   & 14.791$\pm$2.384 & 73.43 &
    \textbf{100302A} & 4.813 & 12.882$\pm$1.483 & 19.697 \\
    100418A & 0.6235 & 0.347$\pm$0.048 & 1.149 &
    100424A & 2.465 & 21.380$\pm$1.477 & 93.262 \\
    100425A & 1.755 & 3.715$\pm$0.684 & 2.819 &
    \textbf{100513A} & 4.772 & 57.544$\pm$3.975 & 35.59 \\
    100615A & 1.398 & 25.704$\pm$0.592 & 41.672 &
    100621A & 0.542 & 16.218  & 3.374 \\
    \textbf{100704A} & 3.6   & 158.489$\pm$3.649 & 142.154 &
    100814A & 1.44  & 48.978$\pm$1.128 & 57.199 \\
    \textbf{100901A} & 1.408 & 10.965$\pm$1.515 & 46.952 &
    \textbf{100902A} & 4.5   & 120.226$\pm$8.305 & 380.507 \\
    \textbf{100906A} & 1.727 & 91.201  & 27.27 &
    101219B & 0.5519 & 1.698$\pm$0.313 & 0.214 \\
    110213A & 1.46  & 33.113$\pm$2.287 & 66.212 &
    110715A & 0.82  & 21.380$\pm$0.492 & 5.496 \\
    110808A & 1.348 & 1.585$\pm$0.328 & 2.295 &
    111008A & 5     & 234.423$\pm$10.796 & 322.219 \\
    111123A & 3.1516 & 158.489$\pm$7.299 & 67.331 &
    \textbf{111209A} & 0.677 & 43.652$\pm$1.005 & 1.427 \\
    111228A & 0.71627 & 11.749$\pm$0.271 & 4.023 &
    111229A & 1.3805 & 1.698$\pm$0.313 & 6.407 \\
    120118B & 2.943 & 34.674$\pm$1.597 & 26.658 &
    120326A & 1.798 & 21.380$\pm$2.461 & 92.65 \\
    120327A & 2.813 & 64.565$\pm$1.487 & 43.779 &
    120404A & 2.876 & 29.512$\pm$2.039 & 28.733 \\
    120422A & 0.28  & 0.045$\pm$0.007 & 0.052 &
    120521C & 6     & 63.096$\pm$5.811 & 46.249 \\
    120712A & 4.1745 & 60.256$\pm$2.775 & 65.143 &
    \textbf{120802A} & 3.796 & 54.954$\pm$7.592 & 38.096 \\
    120811C & 2.671 & 48.978$\pm$4.511 & 38.44 &
    120922A & 3.1   & 128.825$\pm$14.832 & 43.932 \\
    \textbf{121024A} & 2.298 & 13.804$\pm$1.271 & 19.42 &
    \textbf{121128A} & 2.2   & 81.283$\pm$3.743 & 46.254 \\
    \textbf{121211A} & 1.023 & 3.631$\pm$0.669 & 3.062 &
    130131B & 2.539 & 5.129$\pm$0.590 & 10.444 \\
    130408A & 3.758 & 66.069$\pm$10.649 & 108.999 &
    130420A & 1.297 & 31.623$\pm$1.456 & 4.377\\
    130511A & 1.3033 & 1.000$\pm$0.161 & 1.181 &
    \textbf{130514A} & 3.6   & 245.471$\pm$5.652 & 55.304 \\
    130603B & 0.3564 & 0.204$\pm$0.009 & 0.277 &
    \textbf{130606A} & 5.913 & 165.959$\pm$11.464 & 190.4\\
    130612A & 2.006 & 2.291$\pm$0.475 & 2.333 &
    \textbf{131030A} & 1.293 & 128.825  & 27.568\\
    \textbf{131103A} & 0.599 & 0.776$\pm$0.089 & 0.651 &
    131105A & 1.686 & 51.286$\pm$3.543 & 19.912 \\
    \textbf{140114A} & 3     & 63.096$\pm$1.453 & 38.2 &
    \textbf{140206A} & 2.73  & 269.153$\pm$6.197 & 214.639\\
    140213A & 1.2076 & 46.774  & 27.792 &
    \textbf{140304A} & 5.283 & 57.544$\pm$3.975 & 455.163\\
    \textbf{140430A} & 1.6   & 7.244$\pm$1.168 & 6.095 &
    \textbf{140512A} & 0.725 & 19.498$\pm$0.449 & 12.788 \\
    140518A & 4.707 & 40.738$\pm$3.752 & 58.399 &
    140614A & 4.233 & 44.668$\pm$12.342 & 35.379 \\
    140629A & 2.275 & 30.200$\pm$2.086 & 28.524 &
    140703A & 3.14  & 83.176$\pm$5.746 & 153.816 \\
    140903A & 0.351 & 0.044$\pm$0.003 & 0.252 &
    141004A & 0.57  & 0.575$\pm$0.027 & 0.286 \\
    141026A & 3.35  & 30.903$\pm$2.135 & 44.513&
    \textbf{141121A} & 1.47  & 30.200$\pm$2.086 & 21.023 \\
    150323A & 0.539 & 5.623$\pm$0.129 & 0.378 &
    150403A & 2.06  & 177.828$\pm$4.095 & 452.607 \\
    150423A & 1.394 & 0.324$\pm$0.045 & 0.456 &
    150424A & 3     & 29.512$\pm$2.039 & 30.343\\
    150910A & 1.359 & 23.442$\pm$1.619 & 62.048 &
    151027A & 0.81  & 13.804$\pm$0.318 & 30.039 \\
    151027B & 4.063 & 47.863$\pm$8.817 & 58.131 &
    151112A & 4.1   & 30.903$\pm$3.558 & 104.129\\
    151215A & 2.59  & 4.786$\pm$0.992 & 9.663 &
    160121A & 1.96  & 5.888$\pm$0.407 & 13.53 \\
    160227A & 2.38  & 41.687$\pm$2.880 & 67.44 &
    160303A & 2.3   & 1.905$\pm$0.351 & 2.446 \\
    160314A & 0.726 & 0.398$\pm$0.055 & 0.585 &
    160327A & 4.99  & 61.660$\pm$4.259 & 84.611 \\
    160804A & 0.736 & 16.596$\pm$0.382 & 0.89 &
    161108A & 1.159 & 3.981$\pm$0.367 & 3.501\\
    161117A & 1.549 & 125.893  & 18.466 &
    170113A & 1.968 & 6.457$\pm$0.595 & 29.68 \\
    170202A & 3.65  & 89.125$\pm$2.052 & 70.381 &
    170519A & 0.818 & 1.995$\pm$0.322 & 2.568\\
    170607A & 0.557 & 6.166$\pm$0.284 & 1.916 &
    170705A & 2.01  & 95.499$\pm$2.199 & 86.809 \\
    170714A & 0.793 & 4.677$\pm$0.431 & 129.891 &
    171205A & 0.0368 & 0.011$\pm$0.001 & 0.003 \\
    171222A & 2.409 & 26.303$\pm$2.423 & 19.17 &
    180115A & 2.487 & 10.965$\pm$1.515 & 16.69 \\
    180325A & 2.25  & 79.433$\pm$1.829 & 135.483 &
    180329B & 1.998 & 32.359$\pm$2.980 & 14.349 \\
    180404A & 1     & 3.467$\pm$0.240 & 1.262 &
    180720B & 0.654 & 97.724$\pm$2.250 & 73.881 \\
\bottomrule																																																																		\end{longtable*}
\clearpage

\begin{longtable*}[ht]{cccc||cccc}
  \caption{Data of GRBs with X-ray flares.}\\
  \toprule
    GRB   & $z$  & $E_{\rm \gamma,iso}~(10^{51}~\rm erg)$ & $E_{\rm flare}~(10^{50}~\rm erg)$&GRB  & $z$  & $E_{\rm \gamma,iso}~(10^{51}~\rm erg)$ & $E_{\rm flare}~(10^{50}~\rm erg)$ \\
  \midrule
    050714B & 2.4383 & 49.900  & 0.556  &
    050406 & 2.44  & 2.300$\pm$0.800 & 1.740\\
    \textbf{050416A} & 0.65  & 0.617  & 0.363  &
    050502B & 5.2   & 26.600  & 41.466  \\
    050607 & 4     & 123.000  & 2.869  &
    050724 & 0.257 & 0.090$\pm$0.020 & 0.210  \\
    \textbf{050730} & 3.97  & 74.131$\pm$5.121 & 43.620  &
    \textbf{050802} & 1.71  & 56.600  & 1.411 \\
    \textbf{050803} & 0.422 & 0.977$\pm$0.068 & 3.800  &
    \textbf{050814} & 5.3   & 99.000  & 4.143  \\
    050819 & 2.5043 & 102.000  & 0.402  &
    050820A & 2.612 & 1033.600$\pm$36.000 & 482.000  \\
    050822 & 1.43  & 25.500  & 0.420  &
    050904 & 6.1   & 1333.6$\pm$138.9 & 370.800  \\
    050908 & 3.35  & 19.700$\pm$3.200 & 3.360  &
    050915A & 2.5273 & 18.0$\pm$13 & 2.960  \\
    050922B & 4.5   & 67.610  & 13.246  &
    051006 & 1.059 & 4.898  & 0.159  \\
    \textbf{051016B} & 0.94  & 0.370  & 0.404  &
    051227 & 0.714 & 1.200  & 0.026  \\
    \textbf{060108} & 2.03  & 5.900  & 0.692  &
    060111A & 2.32  & 14.790  & 13.048  \\
    \textbf{060115} & 3.53  & 44.668$\pm$4.114 & 4.200  &
    060124 & 2.3   & 437.9$\pm$63.9 & 554.900  \\
    \textbf{060210} & 3.91  & 234.423$\pm$10.796 & 69.000  &
    060223A & 4.41  & 97.3$\pm$7.2 & 7.500  \\
    060418 & 1.49  & 135.5$\pm$27.1 & 20.100  &
    060510B & 4.9   & 367.000$\pm$28.700 & 290.800  \\
    060512 & 0.4428 & 0.325  & 0.090  &
    \textbf{060522} & 5.11  & 52.481$\pm$4.834 & 5.320  \\
    \textbf{060526} & 3.21  & 28.184$\pm$3.245 & 221.110  &
    \textbf{060604} & 2.1357 & 4.467$\pm$1.029 & 24.830  \\
    \textbf{060607A} & 3.082 & 52.481$\pm$2.417 & 84.890  &
    \textbf{060707} & 3.43  & 39.811$\pm$3.667 & 13.560  \\
    \textbf{060714} & 2.71  & 47.863$\pm$2.204 & 51.330  &
    060719 & 1.532 & 14.000$\pm$1.300 & 0.340  \\
    \textbf{060729} & 0.54  & 16.000  & 6.783  &
    \textbf{060814} & 0.84  & 138.040  & 5.563  \\
    060904B & 0.703 & 3.640$\pm$0.740 & 15.900  &
    \textbf{060906} & 3.69  & 52.480  & 4.471  \\
    \textbf{060908} & 1.8836 & 72.000  & 0.757  &
    060926 & 3.208 & 11.500  & 5.800  \\
    \textbf{061121} & 1.314 & 63.096$\pm$1.453 & 19.700  &
    070103 & 2.6208 & 10.900  & 0.560  \\
    \textbf{070129} & 2.3384 & 38.905$\pm$3.583 & 175.110  &
    \textbf{070306} & 1.4959 & 82.600  & 11.566  \\
    070318 & 0.836 & 9$\pm$2   & 1.460  &
    \textbf{070721B} & 3.626 & 97.724$\pm$4.500 & 39.740  \\
    070724A & 0.457 & 0.016$\pm$0.003 & 0.290  &
    071021 & 2.452 & 81.800$\pm$8.200 & 22.820  \\
    071031 & 2.692 & 39$\pm$6  & 167.800  &
    071112C & 0.823 & 11.800$\pm$1.900 & 0.720  \\
    071122 & 1.14  & 4.620  & 0.130  &
    080210 & 2.641 & 51.3$\pm$2.13 & 9.110  \\
    \textbf{080310} & 2.4266 & 32.359$\pm$2.980 & 212.600  &
    080607 & 3.036 & 2171.000$\pm$60.000 & 120.000  \\
    080805 & 1.505 & 71.600$\pm$19.000 & 10.600  &
    \textbf{080810} & 3.35  & 109.648$\pm$5.049 & 48.800  \\
    080906 & 2     & 212.000$\pm$12.000 & 36.050  &
    080913 & 6.44  & 86.000$\pm$25.000 & 63.090  \\
    080928 & 1.692 & 28.2$\pm$11.7 & 70.540  &
    \textbf{081008} & 1.9685 & 41.687$\pm$1.920 & 9.780  \\
    \textbf{090407} & 1.4485 & 6.026$\pm$0.971 & 10.700  &
    090417B & 0.35  & 1.570  & 4.159  \\
    \textbf{090423} & 8     & 51.286$\pm$3.543 & 18.100  &
    \textbf{090516} & 3.9   & 295.121$\pm$20.386 & 428.340  \\
    090715B & 3     & 205.000$\pm$19.000 & 246.980  &
    090809 & 2.737 & 18.800$\pm$2.600 & 58.310  \\
    090812 & 2.452 & 271.800$\pm$9.700 & 49.700  &
    \textbf{091029} & 2.752 & 41.687$\pm$1.920 & 1.380  \\
    \textbf{100302A} & 4.813 & 12.882$\pm$1.483 & 118.750  &
    \textbf{100513A} & 4.772 & 57.544$\pm$3.975 & 12.400  \\
    \textbf{100704A} & 3.6   & 158.489$\pm$3.649 & 525.100  &
    100728A & 1.567 & 1140.000$\pm$68.000 & 167.800  \\
    100816A & 0.8034 & 73.000$\pm$0.200 & 1.680  &
    \textbf{100901A} & 1.408 & 10.965$\pm$1.515 & 55.850  \\
    \textbf{100902A} & 4.5   & 120.226$\pm$8.305 & 985.000  &
    \textbf{100906A} & 1.727 & 91.201  & 225.000  \\
    110205A & 1.98  & 483.000$\pm$64.000 & 159.610  &
    110801A & 1.858 & 109.000$\pm$27.000 & 99.340  \\
    111107A & 2.893 & 37.600$\pm$5.500 & 2.610  &
    \textbf{111209A} & 0.677 & 43.652$\pm$1.005 & 90.190  \\
    \textbf{120802A} & 3.796 & 54.954$\pm$7.592 & 27.500  &
    \textbf{121024A} & 2.298 & 13.804$\pm$1.271 & 4.870  \\
    \textbf{121128A} & 2.2   & 81.283$\pm$3.743 & 5.290  &
    \textbf{121211A} & 1.023 & 3.631$\pm$0.669 & 29.400  \\
    121229A & 2.707 & 37.000$\pm$11.000 & 36.200  &
    130427B & 2.78  & 31.6$\pm$17.5 & 4.360  \\
    \textbf{130514A} & 3.6   & 245.471$\pm$5.652 & 325.200  &
    \textbf{130606A} & 5.91  & 165.959$\pm$11.464 & 277.800  \\
    130925A & 0.347 & 150.000$\pm$3.000 & 17.770  &
    131004A & 0.717 & 0.690$\pm$0.030 & 0.110  \\
    \textbf{131030A} & 1.293 & 128.825  & 224.000  &
    \textbf{131103A} & 0.599 & 0.776$\pm$0.089 & 7.510  \\
    131117A & 4.042 & 10.300$\pm$1.800 & 6.190  &
    \textbf{140114A} & 3     & 63.096$\pm$1.453 & 402.100  \\
    \textbf{140206A} & 2.73  & 269.153$\pm$6.197 & 275.000  &
    140301A & 1.416 & 9.500$\pm$1.800 & 1.390  \\
    \textbf{140304A} & 5.283 & 57.544$\pm$3.975 & 161.900  &
    \textbf{140430A} & 1.6   & 7.244$\pm$1.168 & 34.470  \\
    140506A & 0.889 & 14.000$\pm$1.400 & 67.330  &
    \textbf{140512A} & 0.725 & 19.498$\pm$0.449 & 8.040  \\
    140515A & 6.32  & 53.800$\pm$5.800 & 56.300  &
    140710A & 0.558 & 0.257  & 0.150  \\
    \textbf{141221A} & 1.452 & 24.600$\pm$0.350 & 3.850  &
    150206A & 2.087 & 619.000$\pm$45.000 & 346.000  \\
\bottomrule																																																																		\end{longtable*}
\clearpage

\begin{table*}
  \centering
  \caption{Results of the linear regression analyses for GRB samples.}
    \begin{tabular}{cccc}
  \hline
  \hline
    Correlations   & Slope & Intercept & Pearson's $r$ \\
  \hline
  $(E_{\rm plateau}-E_{\rm \gamma,iso})^\dag$ &  0.95$\pm$0.04  &  1.40$\pm$1.96  &  0.89\\
  $(E_{\rm flare}-E_{\rm \gamma,iso})^\ddag$  &  0.80$\pm$0.09  &  9.22$\pm$4.50  &  0.67\\
  $(E_{\rm plateau}-E_{\rm \gamma,iso})^\S$ & 0.85$\pm$0.09 &  7.01$\pm$4.58  &  0.81\\
  $(E_{\rm flare}-E_{\rm \gamma,iso})^\S$  &  0.76$\pm$0.14 &  11.70$\pm$7.43  &  0.61\\
  $[(E_{\rm plateau}+E_{\rm flare})-E_{\rm \gamma,iso}]^\S$  &0.82$\pm$0.08  &  9.08$\pm$4.32  &  0.82\\
  \hline
  $\dag$ The plateau sample.\\
  $\ddag$ The flare sample.\\
  $\S$ The overlapping sample.\\
    \end{tabular}
  \label{tab:addlabel}
\end{table*}
\clearpage

\end{document}